\documentclass[
superscriptaddress,
twocolumn,
amsmath,amssymb,
aps,
prl,
]{revtex4-2}

\usepackage{graphicx,framed}
\usepackage{dcolumn}
\usepackage{bm}
\usepackage{braket}
\usepackage{dsfont}
\usepackage{amsthm}

\newcommand{\bOmega}{\boldsymbol{\Omega}}
\newcommand{\bu}{\boldsymbol{u}}
\newcommand{\br}{\boldsymbol{r}}
\newcommand{\be}{\boldsymbol{e}}
\newcommand{\bR}{\boldsymbol{R}}
\newcommand{\bq}{\boldsymbol{q}}
\newcommand{\bk}{\boldsymbol{k}}
\newcommand{\bB}{\boldsymbol{B}}

\newcommand{\mU}{\mathcal{U}}
\newcommand{\mW}{\mathcal{W}}
\newcommand{\bt}{\boldsymbol{t}}

\newcommand{\amp}{\mathcal{A}}

\begin{document}

\preprint{}

\title{Gyromagnetic Quantum Friction in Rayleigh Vorticity Baths}
\author{Mamoru Matsuo}
\email{mamoru@ucas.ac.cn}
\affiliation{
Kavli Institute for Theoretical Sciences, University of Chinese Academy of Sciences, Beijing, 100190, China
}
\affiliation{
CAS Center for Excellence in Topological Quantum Computation, University of Chinese Academy of Sciences, Beijing 100190, China
}
\affiliation{
Advanced Science Research Center, Japan Atomic Energy Agency, Tokai, 319-1195, Japan
}
\affiliation{
RIKEN Center for Emergent Matter Science (CEMS), Wako, Saitama 351-0198, Japan
}

\author{Ryotaro Sano}
\affiliation{Institute for Solid State Physics, The University of Tokyo, Kashiwa 277-8581, Japan}

\author{Ai Yamakage}
\affiliation{Department of Physics, Nagoya University, Nagoya 464-8602, Japan}
\author{Hiroshi Funaki}
\affiliation{
Center for Spintronics Research Network, Keio University, Yokohama 223-8522, Japan
}
\affiliation{
Kavli Institute for Theoretical Sciences, University of Chinese Academy of Sciences, Beijing, 100190, China
}

\author{Tatsuhiko N. Ikeda}
\affiliation{Faculty of Social Informatics, ZEN University, Zushi, Kanagawa, 249-0007, Japan}
\affiliation{RIKEN Center for Quantum Computing, Wako, Saitama 351-0198, Japan}
\affiliation{Department of Applied Physics, Hokkaido University, Sapporo, Hokkaido, 060-8628, Japan}

\begin{abstract}
We identify an intrinsic zero-temperature relaxation channel for near-surface spins gyromagnetically coupled to Rayleigh-wave vorticity. This surface-mode contribution requires no thermal phonons, unlike Raman relaxation, and is fixed by Rayleigh vorticity rather than material-specific $g$-factor modulation. The Rayleigh-vorticity bath is super-Ohmic and evanescent with depth, producing field and depth scalings of spin relaxation. These scalings establish shallow spin sensors and hybrid surface-acoustic-wave spin interfaces as detectors of Rayleigh-wave acoustic quantum friction in solids.
\end{abstract}

\maketitle

\paragraph{Introduction---}
Quantum control and readout of individual spins in solids are central to quantum sensing and quantum information science~\cite{degen2017Rev.Mod.Phys.,wolfowicz2021NatRevMater}. Near-surface spins, such as nitrogen-vacancy centers in diamond~\cite{Jelezko2006,Myers2014,Romach2015}, donor-bound electrons in semiconductors~\cite{Tyryshkin2012}, single-electron quantum dots near interfaces~\cite{Vandersypen2017,Khaetskii2001}, and point defects in two-dimensional materials~\cite{vaidya2023quantum,nie2023surface}, provide direct access to localized quantum degrees of freedom while remaining in close proximity to mechanical modes. The performance of such devices is constrained by both the longitudinal relaxation time $T_1$ and the transverse coherence time $T_2$. In the present work, we focus on intrinsic energy relaxation because it sets a nonzero low-temperature contribution to $T_1^{-1}$ and hence an intrinsic upper limit on $T_1$; when additional low-frequency dephasing is weak, it fixes the corresponding lower bound on $T_2^{-1}$. For spins located a few nanometers from a traction-free surface, elastic boundary conditions reshape phonon eigenmodes and hence can qualitatively modify relaxation pathways relative to the bulk. Identifying the dominant disorder-independent near-surface channel is therefore essential.

Surface acoustic waves (SAWs) provide coherent phonons in the radio-to-gigahertz range and form a controllable interface between localized degrees of freedom and elastic motion~\cite{Barnes2000PRB,Gustafsson2012,Martin2014,Shuetz2015,Riedinger2018,Delsing_2019,Qiao2023Science}. SAW coupling has been studied for localized spins~\cite{Golter2016PRL,Golter2016PRX,Vandersypen2017,Whiteley2019,Xu2020SciAdv,Maity2020NatCommun,Hwang2024PRL,Matsumoto2024,Liao2024PRL,Chen2025PRL}, for flying electrons~\cite{Naber2006PRL,McNeil2011,Hermelin2011,Yamamoto2012,Takada2019,Chen2015,Bertrand2016,Ito2021PRL,Edlbauer2021,Jadot2021,Wang2022PRX,Wang2023,Shaju2025,Wang2025JPhysD}, and for superconducting qubits~\cite{Manenti2017,Yiwen2017,Satzinger2018,Bienfait2019,Bienfait2020PRX,Kitzman2023}. Rayleigh-type SAWs are especially relevant here because their elliptical particle motion carries a finite lattice vorticity confined near the surface; the zero-point vorticity is therefore an unavoidable part of the environment seen by shallow spins. This zero-temperature near-surface channel is experimentally accessible because single-spin detection and readout now permit direct access to relaxation dynamics by complementary techniques, including magnetic resonance force microscopy~\cite{rugar2004Nature}, torque-based detection~\cite{losby2015Science,burgess2013Science,kim2016NatCommun}, scanning tunneling microscopy electron spin resonance~\cite{baumann2015Science}, and optical spin-noise measurements~\cite{oestreich2005Phys.Rev.Lett.,atature2007NaturePhys,Gundin2025PRL}. In nanomechanics, torque detection of spin-angular-momentum transfer has reached sensitivities where intrinsic near-surface relaxation-rate limits become experimentally relevant~\cite{Wallis-APL-2006-09,Zolfagharkhani2008,Harii-2019-NatCommun,Mori2020}.

\begin{figure}
    \centering
    \includegraphics[width=\columnwidth]{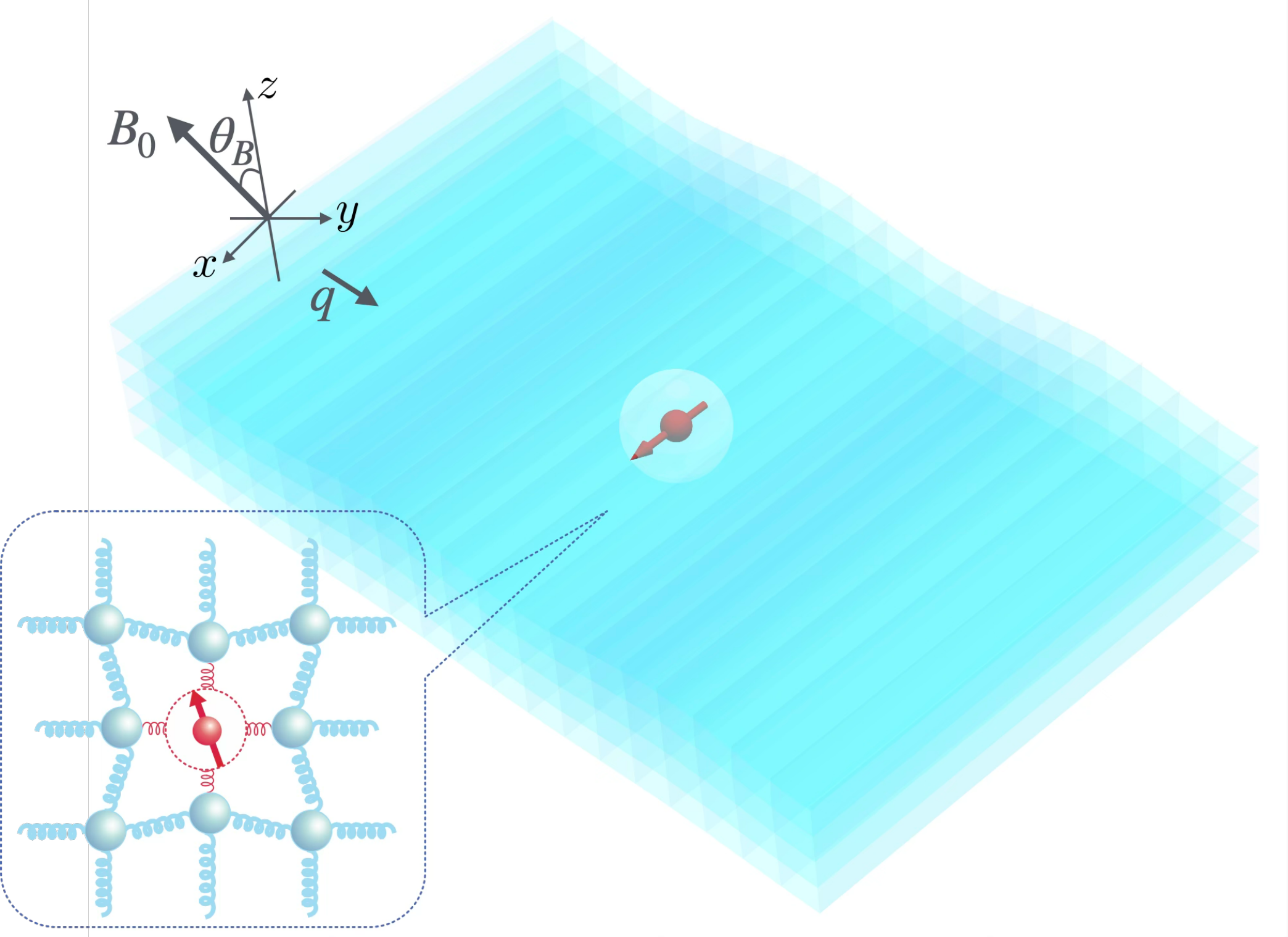}
    \caption{
A near-surface spin-$1/2$ is located at a depth $d$ below a traction-free surface ($z=0$) of an elastic half-space.
The spin couples linearly to the local lattice vorticity generated by Rayleigh surface acoustic waves with in-plane wave vector $\bq$.
A static magnetic field $\bB_0$ defines the quantization axis, tilted by an angle $\theta_B$ from the surface normal $\hat{z}$.
}
    \label{fig:1}
\end{figure}

Gyromagnetism, namely the response of spin to mechanical rotation, provides a microscopic mechanism for such a channel~\cite{Barnett1915,EdH1915,barnett1935Gyromagnetic,scott1962review}. Through the Barnett-type spin-vorticity interaction, a spin couples linearly to the local lattice vorticity~\cite{Matsuo2013,Kobayashi2017,kurimune2020Highly,tateno2020Electrical,kurimune2020Observation,tateno2021Einstein,takahashi2016Spin,takahashi2020Giant,tabaeikazerooni2020Electron,tabaeikazerooni2021Electrical,tokoro2022spin}. For a spin-$1/2$ doublet this coupling is direct, because it is linear in spin, whereas conventional magnetoelastic couplings associated with quadratic spin operators are absent or strongly reduced. Spin--vorticity physics has been discussed across widely separated frequency scales, including rigid-body rotation and liquid-metal flow~\cite{chudo2014Observation,chudo2015Rotational,harii2015Line,chudo2021Barnett,chudo2025mechanical,ono2015Barnett,ogata2017Gyroscopic,hirohata2018magneto,imai2018observation,imai2019angular,wood2017magnetic,wood2018Sci.Adv.,jin2024NatCommun,takahashi2016Spin,takahashi2020Giant,tabaeikazerooni2020Electron,tabaeikazerooni2021Electrical,tokoro2022spin}, gigahertz SAWs~\cite{Kobayashi2017,kurimune2020Highly,tateno2020Electrical,kurimune2020Observation,tateno2021Einstein}, and extreme vorticity fields inferred in heavy-ion collisions~\cite{adamczyk2017global,adam2018global,adam2019polarization,acharya2020evidence,adam2021global}. However, it remains unresolved whether the zero-point vorticity bath of Rayleigh modes provides a symmetry-allowed and quantitatively relevant gyromagnetic relaxation pathway for an individual near-surface spin, and, if present, what universal scaling it imposes in frequency, depth, field orientation, and temperature.

In this Letter, we identify such a gyromagnetic relaxation process for a near-surface spin coupled to a Rayleigh-vorticity bath. Zero-point fluctuations of the Rayleigh vorticity induce spin flips and yield a finite longitudinal relaxation rate in the limit $T\to0$, in sharp contrast to Raman-type two-phonon processes, which require thermally occupied phonons and are therefore suppressed at low temperature. We show that the associated Rayleigh-vorticity spectral density is super-Ohmic and evanescent with depth, leading to characteristic dependences on frequency, depth, and field orientation and hence to a geometry-controlled lower bound on $T_1^{-1}$. At the same time, the same super-Ohmic structure clarifies the implication of this intrinsic channel for $T_2$ by suppressing Markovian pure dephasing in the absence of additional low-frequency noise. This vacuum-persistent dissipation constitutes an acoustic form of quantum friction for a localized spin~\cite{pendry1997,volokitin2007Rev.Mod.Phys.,manjavacas2010Phys.Rev.Lett.}.

\paragraph{Setup---}
We consider a single near-surface spin $S\!=\!1/2$ gyromagnetically coupled to the vorticity fields of Rayleigh SAW and of bulk transverse phonons in an isotropic elastic half-space $z<0$ with a traction-free surface at $z=0$ and outward normal $\hat{\boldsymbol z}$~\cite{Landau1986}. A single spin is located at $\br_0=(\bR_0,-d)$ below the surface, and a static magnetic field $\boldsymbol B_0=B_0\hat{\boldsymbol n}$ sets the spin quantization axis $\hat{\boldsymbol n}$ (Fig.~\ref{fig:1}). 
The total Hamiltonian reads
\begin{align}
\mathcal H=\mathcal H_{\text{el}}+\mathcal H_{\text{spin}}+\mathcal H_{\text{int}},
\label{eq:Htot}
\end{align}
with elastic part
\begin{align}
\mathcal H_{\text{el}}&=\int_{z<0}d^3 r\,
\Bigl[\frac{\rho}{2}\,|\dot{\bu}|^2
+\frac{\lambda}{2}\,(\nabla\!\cdot\!\bu)^2
+\mu\,u_{\alpha\beta}u_{\alpha\beta}\Bigr],
\label{eq:Hel}
\end{align}
where $u_{\alpha\beta}=\tfrac12(\partial_\alpha u_\beta+\partial_\beta u_\alpha)$ denotes the symmetric strain tensor (Einstein summation implied), $\rho$ is the mass density, and $(\lambda,\mu)$ are the Lamé coefficients~\cite{Landau1986}. 
For simplicity, we take the quantization axis along $\hat{\boldsymbol z}$ (i.e., $\hat{\bm{n}}=\hat{\bm{z}}$ for the moment). The Zeeman energy and the Barnett-type spin-vorticity interaction~\cite{Matsuo2013} are given by
\begin{equation}
    \mathcal H_{\text{spin}}=\frac{\hbar\omega_0}{2}\sigma_z, \quad
    \mathcal H_{\text{int}}=-\frac{\hbar}{2}\boldsymbol{\sigma}\cdot\boldsymbol{\Omega}(\br_0,t),
    \label{eq:Hamiltonians}
\end{equation}
where $\omega_0=\gamma B_0$ is the Larmor frequency and $\boldsymbol{\Omega}=\frac{1}{2}\nabla\times\dot{\bm{u}}$ is the local lattice vorticity field.

Crucially, this gyromagnetic interaction represents a fundamental coupling channel for $S=1/2$ systems. Unlike conventional magnetoelastic mechanisms that rely on quadratic spin operators ($\propto S_i S_j$) associated with single-ion anisotropy, which reduce to a constant for a spin-1/2 doublet due to the vanishing quadrupole moment, the vorticity coupling is linear in $\boldsymbol{\sigma}$. Consequently, it serves as a dominant and universal pathway for spin-lattice relaxation in two-level systems where quadrupolar strain couplings are forbidden.

Let us quantize the elastic field in the bulk and surface normal modes. In the bulk, eigenmodes are longitudinal ($L$) and transverse ($T$) plane waves characterized by three-dimensional wave vector $\bm{k}$ with dispersion relations $\omega_{\bk,L/T}=c_{L/T} k$  ($k\equiv|\bm{k}|$). Here the sound velocities are $c_L=\sqrt{(\lambda+2\mu)/\rho}$ and $c_T=\sqrt{\mu/\rho}$, and their ratio depends only on the Poisson ratio $\nu =\lambda/[2(\lambda+\mu)]$. Writing the volume as $V$ and the polarization vectors $\be_{\bk,p}$ ($p=L,T^{(1)},T^{(2)}$), the bulk displacement field is quantized as
\begin{align}
\bu_{\text{bulk}}(\br,t)=\sum_{\bk,p}
\sqrt{\frac{\hbar}{2\rho V\,\omega_{\bk,p}}}\,
\be_{\bk,p}
\Bigl(b_{\bk,p}\,e^{i(\bk\cdot\br-\omega_{\bk,p} t)}+{\rm h.c.}\Bigr).
\label{eq:u_bulk}
\end{align}
At the surface, there exists a Rayleigh surface branch characterized by two-dimensional wave vectors $\bm{q}$ with dispersion $\omega_{\bq,R}=c_R q$ ($c_R<c_T$, $q\equiv|\bm{q}|$) confined within a depth of order $q^{-1}$. Writing $\br=(\bR,z)$ and normalizing to the surface area $A$~\cite{Nakane2024PRB}, the Rayleigh displacement reads
\begin{align}
\bu_R(\br,t)&=\sum_{\bq}
\sqrt{\frac{\hbar q}{2\rho A\,\omega_{\bq,R}}}\,
\big[\mU_{\bq}(z)\,\hat{\bq}+\mW_{\bq}(z)\,\hat{\boldsymbol z}\big]\notag\\
&\quad\times
\Bigl(r_{\bq}\,e^{i(\bq\cdot\bR-\omega_{\bq,R} t)}+{\rm h.c.}\Bigr),
\label{eq:u_R}
\end{align}
where $\mU_{\bq}(z)$ and $\mW_{\bq}(z)$ are real, dimensionless depth profiles that solve the Rayleigh problem with the traction-free boundary condition at $z=0$ and decay for $z\to-\infty$~\cite{Landau1986}.
Their explicit forms and the canonical normalization are summarized in the Supplemental Material~\footnote{See the Supplemental Materials for details of the quantization procedure.}.

Importantly, longitudinal bulk modes are irrotational and do not contribute to the vorticity ($\bOmega_L=0$), whereas the transverse bulk modes and the Rayleigh surface modes generate finite vorticity. These vorticity fluctuations constitute the intrinsic gyromagnetic bath inducing quantum friction into spin dynamics, as we will show below.

\paragraph{Zero-temperature longitudinal relaxation---} 
Starting from $\mathcal H_{\text{int}}=-(\hbar/2)\,\boldsymbol{\sigma}\cdot\boldsymbol{\Omega}(\br_0,t)$, only the components of $\boldsymbol{\Omega}$ transverse to the quantization axis $\hat{\boldsymbol n}$ induce spin flips. 
Treating the interaction as a weak perturbation and assuming a separation of timescales between the fast bath fluctuations and the slow spin dynamics (the Born-Markov approximation), the downward and upward transition rates are given by Fermi's golden rule: 
$\Gamma_{\downarrow,\uparrow} = S_{\Omega_\perp}(\pm\omega_0)/4$, $S_{\Omega_\perp}(\omega) = J(\omega)\coth\left(\frac{\hbar\omega}{2k_B T}\right)$, 
where $S_{\Omega_\perp}(\omega)$ is the symmetrized noise and $J(\omega)$ is the total transverse vorticity spectral density at the spin location.
In the present geometry this can be written as 
$J(\omega)\equiv J_R(\omega;d)+J_\mathrm{bulk}(\omega)$,
where $J_R(\omega;d)=\kappa_R\,\omega^5\exp[-2\eta\omega d/c_R]$ is the Rayleigh surface contribution with explicit depth dependence and $J_\mathrm{bulk}(\omega)$ is the bulk transverse contribution. As argued below, $J(\omega)$ is dominated by $J_R(\omega;d)$ for shallow spins in the relevant frequency range, allowing us to neglect $J_\mathrm{bulk}(\omega)$ whenever subleading. The longitudinal relaxation rate is then
\begin{equation}
  \frac{1}{T_1}=\Gamma_\downarrow+\Gamma_\uparrow
  =\frac{1}{2}\,J_R(\omega_0;d)\,\coth\left(\frac{\hbar\omega_0}{2k_B T}\right).
  \label{eq:T1-general}
\end{equation}
Taking the zero-temperature limit of Eq.~\eqref{eq:T1-general} immediately shows that this is a genuine quantum-relaxation channel:
$ \lim_{T\to0} T_1{}^{-1}
  =\frac{1}{2}\,J_R(\omega_0;d)\;>\;0$. 
Once the Rayleigh spectral density is specified microscopically, Eq.~\eqref{eq:T1-general} yields a geometry-controlled, nonzero zero-temperature relaxation rate, with the explicit angular and depth dependence obtained below.
This vacuum-persistent character is precisely why we refer to this mechanism as \textit{gyromagnetic quantum friction}.

For completeness we generalize to a tilted Zeeman field $\mathcal H_\mathrm{spin}= \tfrac{\hbar\omega_0}{2}( \cos\theta_B \sigma_z + \sin\theta_B \sigma_x)$. For Rayleigh modes, $\boldsymbol{\Omega}_R$ lies strictly in the surface plane ($\hat{\boldsymbol z}\cdot\boldsymbol{\Omega}_R=0$) and is polarized along $\hat{\boldsymbol t}_{\bq}=\hat{\boldsymbol z}\times\hat{\bq}$. Orientational averaging over in-plane propagation directions then yields
$  \big\langle 1-(\hat{\boldsymbol n}\cdot\hat{\boldsymbol t}_{\bq})^2\big\rangle_{\varphi_{\bq}}
  =\frac{1+\cos^2\theta_B}{2}\equiv f(\theta_B)$,
and with $J(\omega)=J_R(\omega;d)$ one finds
\begin{equation}
  \frac{1}{T_1(\theta_B,d,T)}
  =\frac{f(\theta_B)}{2}\,J_R(\omega_0;d)\,
  \coth\!\Bigl(\frac{\hbar\omega_0}{2k_B T}\Bigr).
  \label{eq:T1-angle}
\end{equation}
It is useful to introduce the Rayleigh-evanescent length
\begin{equation}
  \ell_R=\frac{c_R}{2\eta\omega_0}.
  \label{eq:ell_R}
\end{equation}
The Rayleigh-evanescent length is the decay length of the relaxation rate at the resonant Rayleigh wave vector; since $\omega_0=\gamma B_0$, it decreases inversely with the Zeeman magnetic field. In terms of $\ell_R$, Eq.~\eqref{eq:T1-angle} yields
\begin{equation}
\begin{aligned}
  k_B T\ll\hbar\omega_0:\quad &
  T_1^{-1}\simeq \tfrac12\,f(\theta_B)\,\kappa_R\,\omega_0^{5}e^{-d/\ell_R},\\[2pt]
  k_B T\gg\hbar\omega_0:\quad &
  T_1^{-1}\simeq \frac{f(\theta_B)}{\hbar}\,\kappa_R\,\omega_0^{4}\,k_B T\,e^{-d/\ell_R},
\end{aligned}
\label{eq:T1-limits}
\end{equation}
and the evanescent depth dependence
\begin{equation}
  T_1^{-1}(d)=T_1^{-1}(0)\,e^{-d/\ell_R}.
  \label{eq:T1-depth}
\end{equation}
Thus, the longitudinal relaxation is set entirely by the Rayleigh vorticity bath through $J_R(\omega_0;d)$, with a quantum-friction $T\to0$ limit and a geometry-controlled crossover in frequency, temperature, and depth.

\paragraph{Bulk contribution---}
Bulk transverse phonons provide a second, super-Ohmic relaxation channel by generating vorticity at the spin position. Their contribution to the spectral density is
$  J_T(\omega) = \frac{\hbar}{64\pi}\,\frac{\omega^5}{\rho\,c_T^{5}},$
while at the Zeeman frequency the Rayleigh branch contributes $J_R(\omega_0;d)=\kappa_R\,\omega_0^5\,e^{-d/\ell_R}$. The two channels share the same $\omega^5$ scaling and have different prefactors and depth dependence. Their relative weight at the Zeeman frequency is
\begin{align}
  \frac{J_T(\omega_0)}{J_R(\omega_0;d)}
  &= \frac{2}{\pi\,C_R(\nu)^2}\left(\frac{c_R}{c_T}\right)^5
     \exp\!\left(\frac{d}{\ell_R}\right),
\end{align}
The exponential factor determines the depth trend, whereas the small prefactor fixes the crossover depth. For the diamond benchmark used in the Supplemental Material, $(2/\pi C_R^2)(c_R/c_T)^5\simeq1.4\times10^{-2}$. The Rayleigh branch is therefore larger than the bulk transverse branch by roughly a factor of $70$ at the surface, and the bulk branch overtakes it only for $d\simeq4.3\,\ell_R$. Thus reducing the spin depth below $\ell_R$ enhances the Rayleigh contribution through both the evanescent factor and the small bulk-to-Rayleigh prefactor. All material dependence enters through $(\rho,c_R,c_T,C_R(\nu))$ and the evanescent factor via $d$; in particular, the $\omega^5$ power law, the fixed angular factor $f(\theta_B)$, and the single-exponential depth scaling of the Rayleigh term are controlled purely by geometry and elastic boundary conditions rather than by microscopic surface details.

\paragraph{Gyromagnetic process versus conventional processes---}
We now compare the gyromagnetic process identified here with two conventional spin-lattice relaxation mechanisms discussed in prior work: Raman relaxation~\cite{AbragamBleaney1970,Jarmola2012} and strain-induced modulation of the electronic $g$ factor~\cite{Hasegawa1960,Roth1960,Mozyrsky2002,wolfowicz2021NatRevMater}. This comparison separates the microscopic origin of the rate from its phonon order. The gyromagnetic process is fixed by spin-vorticity coupling and the Rayleigh-vorticity mode profile, whereas the conventional processes are controlled by either thermal occupation or material-specific magnetoelastic tensors.
In the present gyromagnetic mechanism the coupling is linear in the vorticity field,
$\mathcal{H}_{\rm int}\propto \boldsymbol{\sigma}\cdot\boldsymbol{\Omega}$.
In the weak-coupling regime this implies that the longitudinal relaxation is determined by a single-phonon spectral density at the transition frequency, $T_1^{-1}\propto J(\omega_0)\coth(\beta\hbar\omega_0/2)$, and remains finite as $T\to0$ whenever $J(\omega_0)>0$ due to spontaneous emission.
The same spectral density determines the frequency renormalization, $\delta\omega_0\propto \mathcal{P}\!\int_0^\infty d\omega\, J(\omega)\bigl[(\omega_0-\omega)^{-1}+(\omega_0+\omega)^{-1}\bigr]$.
Accordingly, the dissipative and dispersive corrections are constrained by a common function $J(\omega)$. In this sense, the gyromagnetic process is spectrally rigid: once $J(\omega)$ is fixed, both the zero-temperature relaxation rate and the dispersive shift follow without additional low-frequency structure.

The first conventional process is Raman relaxation.
Conventional Raman relaxation arises from interactions that are quadratic in phonon coordinates and requires the exchange of two bath excitations.
At the level of perturbation theory the corresponding rate can be written in the generic convolution form
\begin{align}
\frac{1}{T_1^{\rm Raman}}
\propto
\int_0^\infty d\omega\;
\mathcal{J}(\omega,\omega+\omega_0)\,
n(\omega)\bigl[n(\omega+\omega_0)+1\bigr],
\label{eq:prl_raman_convolution}
\end{align}
where $\mathcal{J}(\omega,\omega')$ is a two-frequency kernel involving products of phonon spectra and vertex factors~\footnote{See Supplemental Material for an explicit form of $\mathcal{J}$ and for the connection to standard Raman kernels.}. The Bose factor $n(\omega)$ enforces absorption; hence $(T_1^{\rm Raman})^{-1}\to 0$ as $T\to0$.
For a low-frequency phonon spectrum obeying $J_{\rm ph}(\omega)\propto \omega^{s}$ and a kernel that is smooth at small $\omega$, one finds $(T_1^{\rm Raman})^{-1}\propto T^{2s+1}$ in the window $\hbar\omega_0\ll k_B T\ll \hbar\omega_c$.
A useful scaling estimate follows by equating this Raman law with the strict $T\to0$ limit of the gyromagnetic rate,
\begin{align}
T_\times
\sim
\frac{\hbar}{k_B}
\Bigl(\frac{J(\omega_0)}{\mathcal{C}_{\rm Raman}}\Bigr)^{\!1/(2s+1)},
\label{eq:prl_crossover}
\end{align}
where $\mathcal{C}_{\rm Raman}$ is a positive constant determined by the Raman kernel~\footnote{An explicit expression for $\mathcal{C}_{\rm Raman}$ is given in the Supplemental Material.}. Equation~(\ref{eq:prl_crossover}) shows that Raman processes are parametrically suppressed at low temperature, whereas the gyromagnetic process persists as a temperature-independent contribution to $T_1^{-1}$ fixed by $J(\omega_0)$.
The temperature $T_\times$ is therefore the gyromagnetic--Raman crossover: for $T\lesssim T_\times$ the zero-point gyromagnetic channel is the leading intrinsic phonon contribution, whereas for $T\gtrsim T_\times$ the thermally activated Raman channel can dominate. For a numerical benchmark, the practical comparison uses the finite-temperature gyromagnetic rate, including its $\coth$ factor, rather than only its strict $T\to0$ limit. With the fitted diamond/NV Raman law for a $3~\mathrm{GHz}$ transition, this finite-temperature equality yields $T_\times\approx13~\mathrm{K}$~\cite{Jarmola2012}. This number is a calibrated temperature-scale benchmark rather than a literal realization of the minimal spin-$1/2$ model\footnote{The finite-temperature benchmark is derived in the Supplemental Material.}.

The second conventional process is strain-induced $g$-factor modulation.
This conventional channel was developed in the spin-lattice-relaxation theory of spin-$1/2$ donors and is used in solid-state spin-qubit estimates~\cite{Hasegawa1960,Roth1960,Mozyrsky2002,wolfowicz2021NatRevMater}. In this mechanism, acoustic strain modulates the electronic $g$ tensor and thereby produces a direct spin-lattice relaxation channel in a static magnetic field.
At the level of power counting this process can resemble the present result, because a Rayleigh strain field produces a zero-point rate with an $\omega_0^5$ scaling and an evanescent depth factor.
The distinction is in the prefactor: the $g$-modulation rate is controlled by the material-specific strain derivative of the $g$ tensor, whereas the gyromagnetic rate is fixed by the spin-vorticity coupling and the Rayleigh-vorticity mode profile.
Using the direct-process estimate of Ref.~\cite{Mozyrsky2002}, diamond/NV-scale GHz parameters yield $\Gamma_g^{(0)}(3~\mathrm{GHz})\simeq 1.2\times10^{-9}(\partial g_\perp/\partial\epsilon)^2~\mathrm{s^{-1}}$, about $2.6\times10^{-2}(\partial g_\perp/\partial\epsilon)^2$ of the Rayleigh-vorticity benchmark.
In giant-$g$ materials, the same interband mixing that enhances the Zeeman factor renormalizes the spin-vorticity coupling~\cite{Matsuo2013Renormalization}; consequently, even an estimate based on InSb parameters with $g\simeq-49$, $\delta g_{\rm band}\simeq-51$, and $G_{\rm eff}\simeq-50$ leaves a rate correction of order $4\times10^{-4}$ apart from geometry factors.
Thus $g$-factor modulation is a legitimate conventional magnetoelastic channel and remains a small material-specific correction to the gyromagnetic process in the benchmarks considered here\footnote{The estimate of the $g$-factor modulation correction is given in the Supplemental Material.}.

The comparison can be summarized by two control parameters, the spin depth relative to the Rayleigh-evanescent length and the temperature relative to the gyromagnetic--Raman crossover. Table~\ref{tab:channel-hierarchy} should be read as a visibility map for the Rayleigh-vorticity channel. For $d\gg\ell_R$, the Rayleigh contribution is exponentially suppressed, and the measured low-temperature rate is set by residual non-Rayleigh channels such as bulk spin-lattice relaxation, defect noise, or surface-induced noise. Strain-induced $g$-factor modulation is not shown as a separate dominant regime because the estimates above make it a smaller material-specific direct-process correction to the gyromagnetic channel.
\begin{table}[t]
\caption{Visibility map for the Rayleigh-vorticity channel. Here $\ell_R=c_R/(2\eta\omega_0)$ is the Rayleigh-evanescent length and $T_\times$ is the gyromagnetic--Raman crossover.}
\label{tab:channel-hierarchy}
\begin{ruledtabular}
\begin{tabular}{lcc}
 & $T\lesssim T_\times$ & $T\gtrsim T_\times$ \\
$d\lesssim \ell_R$ & Gyromagnetic & Thermal phonon \\
$d\gg \ell_R$ & Bulk channels & Thermal phonon
\end{tabular}
\end{ruledtabular}
\end{table}

\paragraph{On experimental realization---}
We now separate candidate platforms from the crossover estimate discussed above.
The minimal spin-$1/2$ limit treated here is realized most directly in donor-bound electrons and single-electron quantum dots near interfaces~\cite{Tyryshkin2012,Vandersypen2017,Khaetskii2001}. Although NV centers are not literal spin-$1/2$ realizations, their experimentally isolated two-level subspaces and mature shallow-spin readout justify their use above as a calibrated benchmark for the temperature scale~\cite{Jelezko2006,Jarmola2012,Myers2014,Romach2015}. The comparison above indicates that the Raman crossover need not lie at a prohibitively low temperature. The harder requirement is to reduce surface-induced relaxation channels. A measured relaxation rate contains the gyromagnetic contribution, Raman relaxation, and uncontrolled surface-induced channels, and at low temperature the last term can mask the zero-point Rayleigh-vorticity contribution. High-quality interfaces, controlled spin depth, and high-sensitivity readout are therefore required for a direct test.

\paragraph{Conclusion---}
We have identified a gyromagnetic relaxation process for near-surface spins coupled to zero-point Rayleigh-wave vorticity. The process produces a nonzero low-temperature contribution to $T_1^{-1}$ and is governed by a single super-Ohmic Rayleigh-vorticity spectral density, which fixes its frequency, depth, and field-angle dependences. This boundary-condition contribution differs from Raman relaxation, which is thermally suppressed at low temperature, and from strain-induced $g$-factor modulation, whose prefactor is set by material-specific magnetoelastic coefficients. The benchmark comparison places the relevant crossover on a kelvin scale, leaving a realistic temperature window for testing the intrinsic channel once surface-induced relaxation channels are controlled. The central experimental task is therefore to suppress and characterize these channels, which can mask the zero-point Rayleigh-vorticity contribution. Together, these results establish zero-point Rayleigh vorticity as a boundary-condition origin of gyromagnetic quantum friction in near-surface spin sensors and hybrid SAW-spin interfaces~\cite{degen2017Rev.Mod.Phys.,wolfowicz2021NatRevMater,Whiteley2019,Maity2020NatCommun}.

\begin{acknowledgments}
M.M. is deeply grateful to T. Uto for valuable comments on related 2D material physics. 
We acknowledge fruitful discussions with H. Chudo and S. Watanabe.
M. M. was supported by the National Natural Science Foundation of China (NSFC) under Grant No. 12374126 and by JSPS KAKENHI for Grants (Nos. 23H01839 and 24H00322) from MEXT, Japan. 
A.~Y. was supported by JSPS KAKENHI Grant Nos.~24H00853 and 25K07224.
T.~N.~I. was supported by JSPS KAKENHI Grant No.~25K07178.
\end{acknowledgments}

\bibliographystyle{apsrev4-2}
\bibliography{refs}

\clearpage
\onecolumngrid
\setcounter{equation}{0}
\setcounter{figure}{0}
\setcounter{table}{0}
\setcounter{section}{0}
\setcounter{subsection}{0}
\renewcommand{\theequation}{S\arabic{equation}}
\renewcommand{\thefigure}{S\arabic{figure}}
\renewcommand{\thetable}{S\arabic{table}}
\renewcommand{\thesection}{S\Roman{section}}
\renewcommand{\thesubsection}{S\Roman{section}.\Alph{subsection}}

\begin{center}
{\large \textbf{Supplemental Material for\\
Gyromagnetic Quantum Friction in Rayleigh Vorticity Baths}}
\end{center}
\section{Elastic Eigenmodes and Quantization}
\label{sec:elastic_quantization}

We consider an isotropic elastic half-space $z<0$ with a traction-free surface at $z=0$, mass density $\rho$, and Lam\'e coefficients $(\lambda,\mu)$~\cite{Landau1986,Achenbach1973}. The displacement field $\bu(\br,t)$ satisfies the elastic wave equation
\begin{equation}
\rho\,\ddot{\bu}
=(\lambda+\mu)\,\nabla(\nabla\!\cdot\!\bu)+\mu\,\nabla^2\bu.
\end{equation}
The longitudinal and transverse bulk sound velocities are
\begin{equation}
c_L=\sqrt{\frac{\lambda+2\mu}{\rho}},
\qquad
c_T=\sqrt{\frac{\mu}{\rho}},
\end{equation}
and their ratio is determined by the Poisson ratio
\begin{equation}
\nu=\frac{\lambda}{2(\lambda+\mu)}.
\end{equation}

\subsection{Bulk modes}

In the bulk, the elastic eigenmodes are plane waves labeled by wavevector $\bk$ and polarization $p=L,T^{(1)},T^{(2)}$ with dispersions
\begin{equation}
\omega_{\bk,L}=c_L k,
\qquad
\omega_{\bk,T}=c_T k,
\qquad
k\equiv |\bk|.
\end{equation}
Writing the quantization volume as $V$ and polarization vectors as $\be_{\bk,p}$, the bulk displacement operator is
\begin{align}
\bu_{\text{bulk}}(\br,t)
&=\sum_{\bk,p}
\sqrt{\frac{\hbar}{2\rho V\,\omega_{\bk,p}}}\,
\be_{\bk,p}
\Bigl(
b_{\bk,p}\,e^{i(\bk\!\cdot\!\br-\omega_{\bk,p} t)}
+b_{\bk,p}^\dagger e^{-i(\bk\!\cdot\!\br-\omega_{\bk,p} t)}
\Bigr),
\label{eq:SM_u_bulk}
\end{align}
where the phonon operators obey
\begin{equation}
[b_{\bk,p},b_{\bk',p'}^\dagger]=\delta_{\bk,\bk'}\delta_{p,p'}.
\end{equation}
Longitudinal modes are irrotational and therefore do not contribute to the vorticity bath, whereas the transverse bulk modes do.

\subsection{Rayleigh surface mode}

For the semi-infinite isotropic half-space ($z<0$) with a traction-free surface at $z=0$, the Rayleigh branch is a surface-bound eigenmode propagating with in-plane wavevector $\bq$ and frequency $\omega_{\bq,R}=c_R q$ where $0<c_R<c_T$ and $q\equiv |\bq|$. It is convenient to parameterize the displacement in the P--SV subspace by scalar potentials~\cite{Landau1986,Achenbach1973},
\begin{equation}
\bu_R(\br,t)=\nabla\Phi(\br,t)+\nabla\times\!\big[\Psi(\br,t)\,\hat{\bt}_{\bq}\big],
\qquad
\hat{\bt}_{\bq}\equiv \hat{\boldsymbol z}\times\hat{\bq}.
\label{eq:Rayleigh_pot_repr}
\end{equation}
For a monochromatic surface wave we write
\begin{align}
\Phi(\br,t) &= -\,i\,b\,e^{\kappa_L z}\,e^{i(\bq\cdot\bR-\omega t)},\\
\Psi(\br,t) &= -\,a\,e^{\kappa_T z}\,e^{i(\bq\cdot\bR-\omega t)},
\end{align}
with decay constants
\begin{equation}
\kappa_L=\sqrt{q^2-\omega^2/c_L^2},
\qquad
\kappa_T=\sqrt{q^2-\omega^2/c_T^2}.
\label{eq:kappa_LT}
\end{equation}

The traction-free boundary condition fixes the amplitude ratio and yields the Rayleigh secular equation,
\begin{equation}
a=-\frac{2q\kappa_L}{q^2+\kappa_T^2}\,b,
\end{equation}
\begin{equation}
(q^2+\kappa_T^2)^2 = 4q^2 \kappa_L \kappa_T.
\end{equation}
Introducing $\omega=c_R q$ and
\begin{equation}
\gamma \equiv \sqrt{1-\frac{c_R^2}{c_L^2}},
\qquad
\eta \equiv \sqrt{1-\frac{c_R^2}{c_T^2}},
\end{equation}
we obtain
\begin{equation}
\kappa_L=q\gamma,
\qquad
\kappa_T=q\eta.
\end{equation}

Combining the potentials into a compact mode profile, we write
\begin{align}
\bu_R(\br,t)
&=\amp_{\bq}\boldsymbol{\Phi}_{\bq}(z)\,e^{i(\bq\cdot\bR-\omega t)},\\
\boldsymbol{\Phi}_{\bq}(z)
&= U_{\bq}(z)\,\hat{\bq}+iW_{\bq}(z)\,\hat{\boldsymbol z},
\end{align}
with
\begin{align}
U_{\bq}(z)
&=e^{\kappa_L z}
-\sqrt{\gamma\eta}\,e^{\kappa_T z},
\label{eq:Uq_def}\\
W_{\bq}(z)
&=-\gamma\,e^{\kappa_L z}
+\sqrt{\gamma/\eta}\,e^{\kappa_T z}.
\label{eq:Wq_def}
\end{align}

Quantizing the Rayleigh branch over surface area $A$, the displacement operator becomes
\begin{align}
\bu_R(\br,t)
&=\sum_{\bq}
\sqrt{\frac{\hbar q\,\Gamma(\nu)}{2\rho A\,\omega_{\bq,R}}}\,
\Bigl(
\bigl[U_{\bq}(z)\,\hat{\bq}+iW_{\bq}(z)\,\hat{\boldsymbol z}\bigr]
r_{\bq}\,e^{i(\bq\cdot\bR-\omega_{\bq,R} t)}
+\text{H.c.}
\Bigr),
\label{eq:SM_u_R_raw}
\end{align}
where
\begin{equation}
[r_{\bq},r_{\bq'}^\dagger]=\delta_{\bq,\bq'}.
\end{equation}
The dimensionless mode-shape factor is
\begin{equation}
\Gamma(\nu)=\frac{2\eta^2\gamma}{(\gamma-\eta)(\gamma-\eta+2\eta^2\gamma)}.
\end{equation}

It is convenient to absorb $\Gamma(\nu)$ into normalized depth profiles
\begin{equation}
\mU_{\bq}(z)\equiv \sqrt{\Gamma(\nu)}\,U_{\bq}(z),
\qquad
\mW_{\bq}(z)\equiv \sqrt{\Gamma(\nu)}\,W_{\bq}(z),
\end{equation}
The normalized profiles put the Rayleigh displacement in the same form used in the main text,
\begin{align}
\bu_R(\br,t)
&=\sum_{\bq}
\sqrt{\frac{\hbar q}{2\rho A\,\omega_{\bq,R}}}\,
\Bigl(
\bigl[\mU_{\bq}(z)\,\hat{\bq}+i\mW_{\bq}(z)\,\hat{\boldsymbol z}\bigr]
r_{\bq}\,e^{i(\bq\cdot\bR-\omega_{\bq,R} t)}
+\text{H.c.}
\Bigr).
\label{eq:SM_u_R}
\end{align}

The profiles $\mU_{\bq}(z)$ and $\mW_{\bq}(z)$ are real, dimensionless, satisfy the traction-free boundary condition at $z=0$, and decay exponentially as $z\to-\infty$.

\subsection{Rayleigh vorticity and the exact geometric prefactor}
\label{subsec:SM_Rayleigh_vorticity}

The relative phase between the in-plane and out-of-plane displacement components in Eq.~\eqref{eq:SM_u_R} is essential. For a Rayleigh mode propagating along $\hat{\bq}$, the vorticity lies along the in-plane transverse direction $\hat{\bt}_{\bq}\equiv \hat{\boldsymbol z}\times \hat{\bq}$, with
\begin{equation}
(\nabla\times \bu_R)\cdot \hat{\bt}_{\bq}
=
\partial_z u_{\bq}- (\hat{\bq}\cdot\nabla)u_z.
\end{equation}
Using Eq.~\eqref{eq:SM_u_R} one obtains the local Rayleigh-vorticity operator
\begin{align}
\boldsymbol{\Omega}_R(\br,t)
&=
-\frac{i}{2}\sum_{\bq}\omega_{\bq,R}
\sqrt{\frac{\hbar q}{2\rho A\,\omega_{\bq,R}}}\,
\bigl[\partial_z \mU_{\bq}(z)+q\,\mW_{\bq}(z)\bigr]
\hat{\bt}_{\bq}\notag\\
&\quad\times
\Bigl(
r_{\bq}\,e^{i(\bq\cdot\bR-\omega_{\bq,R} t)}
-\text{H.c.}
\Bigr).
\label{eq:SM_Omega_R}
\end{align}
The depth derivative simplifies exactly because the longitudinally decaying branch cancels:
\begin{align}
\partial_z U_{\bq}(z)+q\,W_{\bq}(z)
&=
\kappa_L e^{\kappa_L z}
-q\eta\sqrt{\gamma\eta}\,e^{\kappa_T z}
-q\gamma e^{\kappa_L z}
+q\sqrt{\gamma/\eta}\,e^{\kappa_T z}\notag\\
&=
q\sqrt{\gamma/\eta}\,(1-\eta^2)e^{\kappa_T z}.
\label{eq:SM_dU_plus_qW_raw}
\end{align}
After inserting the normalization factor $\Gamma(\nu)$, this becomes
\begin{align}
\partial_z \mU_{\bq}(z)+q\,\mW_{\bq}(z)
&=
q\sqrt{\Gamma(\nu)\gamma/\eta}\,(1-\eta^2)e^{\kappa_T z}\notag\\
&\equiv
\frac{q\,C_R(\nu)}{4}\,e^{\kappa_T z},
\label{eq:SM_dU_plus_qW}
\end{align}
which defines the exact dimensionless Rayleigh geometric factor
\begin{equation}
C_R(\nu)
=
4(1-\eta^2)\sqrt{\Gamma(\nu)\gamma/\eta},
\qquad
C_R(\nu)^2
=
16\,\Gamma(\nu)\,\frac{\gamma}{\eta}(1-\eta^2)^2.
\label{eq:SM_CR_exact}
\end{equation}
Evaluated at the spin position $\br_0=(\bR_0,-d)$, Eq.~\eqref{eq:SM_Omega_R} reduces to
\begin{align}
\boldsymbol{\Omega}_R(\br_0,t)
&=
-\frac{i\,C_R(\nu)}{8}\sum_{\bq}\omega_{\bq,R}q
\sqrt{\frac{\hbar q}{2\rho A\,\omega_{\bq,R}}}\,
e^{-\eta q d}\,
\hat{\bt}_{\bq}\notag\\
&\quad\times
\Bigl(
r_{\bq}\,e^{i(\bq\cdot\bR_0-\omega_{\bq,R} t)}
-\text{H.c.}
\Bigr).
\label{eq:SM_Omega_R_at_spin}
\end{align}
With the spectral-density convention used in the main text,
\begin{equation}
J_R(\omega;d)
=
2\pi\sum_{\bq}|g_{\bq}(d)|^2\,
\delta(\omega-\omega_{\bq,R}),
\label{eq:SM_JR_sum}
\end{equation}
where
\begin{equation}
g_{\bq}(d)
=
\frac{C_R(\nu)}{8}\,\omega_{\bq,R}q
\sqrt{\frac{\hbar q}{2\rho A\,\omega_{\bq,R}}}\,
e^{-\eta q d},
\end{equation}
the two-dimensional mode sum yields
\begin{align}
J_R(\omega;d)
&=
\frac{\hbar\,C_R(\nu)^2}{128\,\rho\,c_R^5}\,
\omega^5 e^{-2\eta \omega d/c_R}\notag\\
&\equiv
\kappa_R\,\omega^5 e^{-2\eta \omega d/c_R},
\qquad
\kappa_R=
\frac{\hbar\,C_R(\nu)^2}{128\,\rho\,c_R^5}.
\label{eq:SM_JR_exact}
\end{align}
Equations~\eqref{eq:SM_CR_exact} and \eqref{eq:SM_JR_exact} provide the closed-form origin of the geometric prefactor used in the main text.

\section{Rayleigh--Bulk Spectral-Weight Ratio}
\label{sec:SM_bulk_rayleigh_ratio}

The comparison between the Rayleigh surface branch and the bulk transverse branch requires the exponential depth factor and the dimensionless prefactor multiplying it. Combining Eq.~\eqref{eq:SM_JR_exact} with the bulk transverse vorticity spectrum used in the main text,
\begin{align}
J_T(\omega)
=
\frac{\hbar}{64\pi}\frac{\omega^5}{\rho c_T^5},
\end{align}
yields, at the Zeeman frequency,
\begin{align}
\frac{J_T(\omega_0)}{J_R(\omega_0;d)}
&=
A_T(\nu)\exp\!\left(\frac{d}{\ell_R}\right),
\label{eq:SM_bulk_rayleigh_ratio}\\
A_T(\nu)
&\equiv
\frac{2}{\pi C_R(\nu)^2}
\left(\frac{c_R}{c_T}\right)^5,
\qquad
\ell_R\equiv\frac{c_R}{2\eta\omega_0}.
\label{eq:SM_bulk_rayleigh_prefactor}
\end{align}
Thus $d/\ell_R$, the depth in units of the Rayleigh-evanescent length, controls the exponential growth of the bulk-to-Rayleigh ratio, while $A_T(\nu)$ fixes the near-surface offset.

For the diamond parameters used below, $C_R^2\simeq25.63$ and $\eta\simeq0.461$. These values yield $c_R/c_T=\sqrt{1-\eta^2}\simeq0.887$ and
\begin{align}
A_T
\simeq
\frac{2}{\pi(25.63)}(0.887)^5
\simeq
1.4\times10^{-2}.
\label{eq:SM_bulk_rayleigh_prefactor_num}
\end{align}
At the surface, therefore, the Rayleigh vorticity spectral density is larger than the bulk transverse one by about $1/A_T\simeq70$. The crossover depth at which the two become equal is
\begin{align}
d_\times^{(T/R)}
=
\ell_R\ln\frac{1}{A_T}
\simeq
4.3\,\ell_R .
\label{eq:SM_bulk_rayleigh_depth_cross}
\end{align}
For the $3~\mathrm{GHz}$ benchmark, $\ell_R\simeq0.63~\mu\mathrm{m}$, giving $d_\times^{(T/R)}\simeq2.7~\mu\mathrm{m}$. This estimate shows that reducing the spin depth strengthens the Rayleigh channel through the evanescence and through the small bulk-to-Rayleigh prefactor.

\section{Raman Kernel and Low-Temperature Prefactor}
\label{sec:SM_raman_kernel}

In this section we derive the two-frequency Raman kernel $\mathcal{J}(\omega,\omega')$ for a generic two-phonon relaxation mechanism. We evaluate the Raman prefactor entering the low-temperature power law quoted in the main text.

\subsection{Quadratic coupling and Raman operator}
\label{subsec:SM_quadratic_coupling}

Consider an effective interaction that induces spin flips through a bath operator that is quadratic in phonon coordinates,
\begin{align}
\mathcal{H}_{\rm int}^{\rm Raman}
=
\sigma_{+}\,\mathcal{B}
+
\sigma_{-}\,\mathcal{B}^{\dagger},
\label{eq:SM_Hint_Raman_def}
\end{align}
where $\mathcal{B}$ is bilinear in bosonic operators. For a harmonic phonon bath one may expand a generic quadratic coordinate coupling in normal modes and retain the number-conserving part relevant for Raman scattering,
\begin{align}
\mathcal{B}
=
\sum_{\lambda,\mu}
\Lambda_{\lambda\mu}\,
b_{\mu}^{\dagger} b_{\lambda}.
\label{eq:SM_B_def}
\end{align}
Here $\lambda,\mu$ label phonon modes with frequencies $\omega_{\lambda},\omega_{\mu}>0$, and $\Lambda_{\lambda\mu}$ collects vertex factors and energy denominators associated with the microscopic origin of the Raman process. Equation~(\ref{eq:SM_B_def}) represents the absorption of a phonon in mode $\lambda$ and emission into mode $\mu$.

\subsection{Two-frequency kernel}
\label{subsec:SM_kernel_definition}

In the weak-coupling regime the downward transition rate is obtained from Fermi's golden rule,
\begin{align}
\Gamma_{\downarrow}^{\rm Raman}
=
\frac{2\pi}{\hbar^{2}}
\sum_{\lambda,\mu}
|\Lambda_{\lambda\mu}|^{2}\,
n(\omega_{\lambda})\bigl[n(\omega_{\mu})+1\bigr]\,
\delta\!\left(\omega_{\mu}-\omega_{\lambda}-\omega_{0}\right),
\label{eq:SM_Gamma_down_sum}
\end{align}
where $n(\omega)=(e^{\beta\hbar\omega}-1)^{-1}$ is the Bose distribution. The convolution structure is made explicit by introducing the two-frequency kernel
\begin{align}
\mathcal{J}(\omega,\omega')
\equiv
2\pi
\sum_{\lambda,\mu}
|\Lambda_{\lambda\mu}|^{2}\,
\delta(\omega-\omega_{\lambda})\,
\delta(\omega'-\omega_{\mu}),
\qquad
\omega,\omega'>0.
\label{eq:SM_J_kernel_def}
\end{align}
Using Eq.~(\ref{eq:SM_J_kernel_def}), Eq.~(\ref{eq:SM_Gamma_down_sum}) becomes
\begin{align}
\Gamma_{\downarrow}^{\rm Raman}
=
\frac{1}{\hbar^{2}}
\int_{0}^{\infty} d\omega
\int_{0}^{\infty} d\omega'\;
\mathcal{J}(\omega,\omega')\,
n(\omega)\bigl[n(\omega')+1\bigr]\,
\delta(\omega'-\omega-\omega_{0}).
\label{eq:SM_Gamma_down_integral2}
\end{align}
Performing the $\omega'$ integration yields
\begin{align}
\Gamma_{\downarrow}^{\rm Raman}
=
\frac{1}{\hbar^{2}}
\int_{0}^{\infty} d\omega\;
\mathcal{J}(\omega,\omega+\omega_{0})\,
n(\omega)\bigl[n(\omega+\omega_{0})+1\bigr].
\label{eq:SM_Gamma_down_integral1}
\end{align}
The upward rate follows analogously and obeys detailed balance; the longitudinal relaxation rate is $T_{1}^{-1}=\Gamma_{\downarrow}+\Gamma_{\uparrow}$. Equation~(\ref{eq:SM_Gamma_down_integral1}) shows that Raman relaxation is thermally activated because the factor $n(\omega)$ enforces phonon absorption. Consequently,
\begin{align}
\lim_{T\to 0}\Gamma_{\downarrow}^{\rm Raman}=0,
\qquad
\lim_{T\to 0}(T_{1}^{\rm Raman})^{-1}=0.
\label{eq:SM_Raman_T0_zero}
\end{align}

\subsection{Connection to standard Raman kernels}
\label{subsec:SM_connection_standard}

The kernel $\mathcal{J}(\omega,\omega')$ reduces to the standard form involving products of one-phonon spectra when the squared vertex factor admits a factorized representation,
\begin{align}
|\Lambda_{\lambda\mu}|^{2}
=
|v_{\lambda}|^{2}\,|v_{\mu}|^{2}\,
\mathcal{K}(\omega_{\lambda},\omega_{\mu}),
\label{eq:SM_vertex_factorization}
\end{align}
with a smooth function $\mathcal{K}$ at low frequency and mode-dependent amplitudes $v_{\lambda}$. Defining the associated one-phonon spectral density,
\begin{align}
J_{\rm ph}(\omega)
\equiv
2\pi\sum_{\lambda}|v_{\lambda}|^{2}\delta(\omega-\omega_{\lambda}),
\qquad
\omega>0,
\label{eq:SM_Jph_def}
\end{align}
Eqs.~(\ref{eq:SM_J_kernel_def}) and (\ref{eq:SM_vertex_factorization}) yield
\begin{align}
\mathcal{J}(\omega,\omega')
=
\frac{1}{2\pi}\,
\mathcal{K}(\omega,\omega')\,
J_{\rm ph}(\omega)\,J_{\rm ph}(\omega').
\label{eq:SM_J_kernel_factorized}
\end{align}
Substituting Eq.~(\ref{eq:SM_J_kernel_factorized}) into Eq.~(\ref{eq:SM_Gamma_down_integral1}) produces the conventional Raman expression as a two-frequency convolution of phonon spectra with a vertex kernel. Different microscopic mechanisms correspond to different low-frequency behavior of $\mathcal{K}(\omega,\omega')$, such as polynomial dependence on $\omega$ and $\omega'$ arising from couplings to strain or strain gradients.

\subsection{Low-temperature prefactor}
\label{subsec:SM_CR_prefactor}

We now evaluate the Raman prefactor under minimal assumptions on the two-frequency kernel. The low-temperature estimate is
\begin{align}
(T_{1}^{\rm Raman})^{-1}
\simeq
\hbar^{-2}\mathcal{C}_{\rm Raman}
\left(\frac{k_{B}T}{\hbar}\right)^{2s+1}.
\label{eq:SM_Raman_lowT_estimate}
\end{align} Starting from Eq.~(\ref{eq:SM_Gamma_down_integral1}), consider the regime
\begin{align}
\hbar\omega_{0}\ll k_{B}T \ll \hbar\omega_{c},
\label{eq:SM_CR_window_def}
\end{align}
where $\omega_{c}$ is a high-frequency cutoff of the phonon spectrum. In this window the integral is dominated by frequencies $\omega\sim k_{B}T/\hbar$, and $\omega_{0}$ may be treated as a small parameter in the kernel. For the factorized form in Eq.~(\ref{eq:SM_J_kernel_factorized}), assume the low-frequency power law
\begin{align}
J_{\rm ph}(\omega)\simeq A\,\omega^{s},
\qquad
\omega\ll \omega_{c},
\label{eq:SM_CR_Jph_power}
\end{align}
and a smooth vertex kernel,
\begin{align}
\mathcal{K}(\omega,\omega')\simeq \mathcal{K}_{0},
\qquad
\omega,\omega'\ll \omega_{c}.
\label{eq:SM_CR_K_smooth}
\end{align}
Then
\begin{align}
\mathcal{J}(\omega,\omega+\omega_{0})
\simeq
\frac{1}{2\pi}\mathcal{K}_{0}A^{2}\,
\omega^{s}(\omega+\omega_{0})^{s}.
\label{eq:SM_CR_J_approx}
\end{align}
Within the window of Eq.~(\ref{eq:SM_CR_window_def}) one may approximate $(\omega+\omega_{0})^{s}\simeq \omega^{s}$, which yields
\begin{align}
\Gamma_{\downarrow}^{\rm Raman}
\simeq
\frac{1}{\hbar^{2}}
\frac{\mathcal{K}_{0}A^{2}}{2\pi}
\int_{0}^{\infty} d\omega\;
\omega^{2s}\,
n(\omega)\bigl[n(\omega)+1\bigr].
\label{eq:SM_CR_Gamma_down_approx}
\end{align}
Using $n(\omega)\bigl[n(\omega)+1\bigr]=e^{\beta\hbar\omega}/(e^{\beta\hbar\omega}-1)^{2}$ and the change of variables $x=\beta\hbar\omega$, Eq.~(\ref{eq:SM_CR_Gamma_down_approx}) becomes
\begin{align}
\Gamma_{\downarrow}^{\rm Raman}
\simeq
\frac{1}{\hbar^{2}}
\frac{\mathcal{K}_{0}A^{2}}{2\pi}
\Bigl(\frac{k_{B}T}{\hbar}\Bigr)^{2s+1}
\int_{0}^{\infty} dx\;
x^{2s}\,
\frac{e^{x}}{(e^{x}-1)^{2}}.
\label{eq:SM_CR_dimensionless}
\end{align}
For $s>1/2$ the integral converges and admits a standard closed form. Expanding
$e^{x}/(e^{x}-1)^{2}=\sum_{m=1}^{\infty} m e^{-mx}$ yields
\begin{align}
\int_{0}^{\infty} dx\;
x^{2s}\,
\frac{e^{x}}{(e^{x}-1)^{2}}
&=
\sum_{m=1}^{\infty} m
\int_{0}^{\infty} dx\;
x^{2s} e^{-mx}
\label{eq:SM_CR_series_step1}\\
&=
\sum_{m=1}^{\infty} m\,
\frac{\Gamma(2s+1)}{m^{2s+1}}
=
\Gamma(2s+1)\sum_{m=1}^{\infty}\frac{1}{m^{2s}}
=
\Gamma(2s+1)\zeta(2s),
\label{eq:SM_CR_series_step2}
\end{align}
where $\Gamma$ is the Gamma function and $\zeta$ is the Riemann zeta function. Substituting Eq.~(\ref{eq:SM_CR_series_step2}) into Eq.~(\ref{eq:SM_CR_dimensionless}) yields
\begin{align}
\Gamma_{\downarrow}^{\rm Raman}
\simeq
\frac{1}{\hbar^{2}}
\mathcal{C}_{\rm Raman}
\Bigl(\frac{k_{B}T}{\hbar}\Bigr)^{2s+1},
\label{eq:SM_CR_final_rate}
\end{align}
with
\begin{align}
\mathcal{C}_{\rm Raman}
\equiv
\frac{\mathcal{K}_{0}A^{2}}{2\pi}\,
\Gamma(2s+1)\zeta(2s).
\label{eq:SM_CR_def}
\end{align}
Equation~(\ref{eq:SM_CR_def}) is the explicit prefactor quoted in the main text. The same scaling applies to $(T_{1}^{\rm Raman})^{-1}$ up to factors of order unity associated with the sum $\Gamma_{\downarrow}+\Gamma_{\uparrow}$ and with the precise definition of the transverse spin operator that couples to $\mathcal{B}$.

\section{Benchmark Estimate of the Crossover Temperature}
\label{sec:SM_crossover_benchmark}

To convert the scaling analysis of the main text into a practical temperature scale, one must compare the finite-temperature gyromagnetic rate with the Raman contribution. The minimal theory developed in the main text applies most directly to genuine spin-$1/2$ systems such as donor-bound electrons and single-electron quantum dots near interfaces~\cite{Tyryshkin2012,Vandersypen2017,Khaetskii2001}. For an order-of-magnitude benchmark, however, it is advantageous to use a platform for which near-surface longitudinal relaxation has already been characterized experimentally over a broad temperature range with high-sensitivity readout. In this respect shallow NV centers in diamond provide the most informative benchmark presently available~\cite{Jelezko2006,Jarmola2012,Myers2014,Romach2015}. The resulting estimate should therefore be interpreted as a calibrated benchmark for the relevant temperature scale, rather than as a literal microscopic realization of the minimal spin-$1/2$ model.

For the Rayleigh gyromagnetic channel,
\begin{align}
\Gamma_{1\phi}(T)
&=
\Gamma_0\,
\coth\!\left(\frac{\hbar\omega_0}{2k_B T}\right),
\label{eq:SM_Gamma1phi_T}\\
\Gamma_0
&=
\frac{1}{2}J_R(\omega_0;d)
=
\frac{\hbar\,C_R(\nu)^2}{256\,\rho\,c_R^5}\,
\omega_0^5 e^{-d/\ell_R},
\qquad
\ell_R=\frac{c_R}{2\eta\omega_0}.
\label{eq:SM_Gamma0_def}
\end{align}
As a representative benchmark we take a few-GHz transition with
\begin{align}
f_0=3.0~\mathrm{GHz},
\qquad
\omega_0=2\pi f_0,
\qquad
d=5~\mathrm{nm},
\qquad
\theta_B=0,
\end{align}
and use the diamond parameters
\begin{align}
\rho=3515.25~\mathrm{kg/m^3},
\qquad
c_R=10.946~\mathrm{km/s},
\qquad
C_R(\nu)^2\simeq 25.63,
\qquad
\eta\simeq 0.461.
\end{align}
Equation~(\ref{eq:SM_Gamma0_def}) then yields
\begin{align}
\Gamma_0 \simeq 4.5\times 10^{-8}~\mathrm{s^{-1}}.
\label{eq:SM_NV_Gamma0_num}
\end{align}

For the Raman side we use the temperature-dependent coefficient extracted by Jarmola \textit{et al.} for NV ensembles in diamond,
\begin{align}
\Gamma_{\rm Raman}(T)=A_5 T^5,
\qquad
A_5 = 2.2\times 10^{-11}~\mathrm{s^{-1}K^{-5}},
\label{eq:SM_NV_Raman_law}
\end{align}
which corresponds to the fitted $T^5$ term in Ref.~\cite{Jarmola2012}. The physically relevant crossover is therefore defined by the finite-temperature equality
\begin{align}
\Gamma_0\,
\coth\!\left(\frac{\hbar\omega_0}{2k_B T_\times}\right)
=
A_5 T_\times^5.
\label{eq:SM_full_crossover}
\end{align}
At $3$ GHz one has
\begin{align}
\frac{\hbar\omega_0}{k_B}\simeq 0.14~\mathrm{K},
\end{align}
This value places the crossover already in the regime $k_B T \gg \hbar\omega_0$ and
\begin{align}
\coth\!\left(\frac{\hbar\omega_0}{2k_B T}\right)
\simeq
\frac{2k_B T}{\hbar\omega_0}.
\end{align}
Equation~(\ref{eq:SM_full_crossover}) then reduces to
\begin{align}
T_\times
\simeq
\left(
\frac{2\Gamma_0 k_B}{A_5\hbar\omega_0}
\right)^{1/4}
\simeq
13~\mathrm{K}.
\label{eq:SM_Tx_NV_num}
\end{align}
For comparison, equating the Raman rate only to the strict $T\to 0$ gyromagnetic rate would yield
\begin{align}
T_\times^{(0)}
=
\left(
\frac{\Gamma_0}{A_5}
\right)^{1/5}
\simeq
4.6~\mathrm{K},
\label{eq:SM_Tx0_NV_num}
\end{align}
which underestimates the true finite-temperature crossover because the gyromagnetic rate itself grows linearly with $T$ once $k_B T \gg \hbar\omega_0$.

This benchmark has a clear interpretation. The relevant temperature scale for the gyromagnetic/Raman crossover need not be prohibitively low: in a well-characterized few-GHz benchmark it lies in the kelvin-to-ten-kelvin range. At the same time, present shallow-diamond experiments exhibit substantial low-temperature surface-induced relaxation backgrounds~\cite{Myers2014,Romach2015}. Accordingly, the estimate in Eq.~(\ref{eq:SM_Tx_NV_num}) is most naturally read as a benchmark for observability rather than as evidence that the zero-point Rayleigh contribution to $T_1^{-1}$ has already been isolated experimentally. Direct observation of the latter will require substantial suppression and characterization of extrinsic near-surface channels.

\section{Comparison with Strain-Induced $g$-Factor Modulation}
\label{sec:SM_g_factor_modulation}

A superficially similar conventional direct spin-lattice channel is obtained when an acoustic strain field modulates the electronic $g$ tensor.  For a spin doublet in a static magnetic field,
\begin{align}
H_Z=\mu_B B_i g_{ij}S_j,
\qquad
\delta g_{ij}=G_{ij,kl}u_{kl},
\end{align}
where $u_{kl}=(\partial_k u_l+\partial_l u_k)/2$ is the strain tensor and $G_{ij,kl}=\partial g_{ij}/\partial u_{kl}$ is a material-specific strain derivative of the $g$ tensor.  Equivalently, in the notation often used for direct spin-lattice relaxation of spin-$1/2$ donors,
\begin{align}
H_g=g\mu_B A_{ijkl}u_{ij}B_kS_l .
\label{eq:SM_H_g_modulation}
\end{align}
This coupling is distinct from the spin-vorticity interaction used in the main text.  Equation~(\ref{eq:SM_H_g_modulation}) is a strain coupling whose magnitude is controlled by the material-dependent tensor $G_{ij,kl}$, whereas the spin-vorticity coupling is the kinematic coupling of spin to the local angular velocity of the lattice.

For an isotropic bulk acoustic bath, the direct-process estimate used in the spin-qubit literature yields~\cite{Mozyrsky2002,Hasegawa1960,Roth1960,wolfowicz2021NatRevMater}
\begin{align}
\Gamma_g(T)
&=
\frac{\hbar\omega_0^5}{2\pi g^2\rho}
\left(\frac{\partial g_\perp}{\partial\epsilon}\right)^2
\left(
\frac{1}{10c_T^5}+\frac{1}{15c_L^5}
\right)
\coth\!\left(\frac{\hbar\omega_0}{2k_BT}\right),
\label{eq:SM_bulk_g_rate}
\end{align}
where $\partial g_\perp/\partial\epsilon$ denotes the transverse strain derivative that mixes the two Zeeman levels.  This expression has the same Bose factor as any direct spin-lattice process and therefore contains a zero-point contribution.  The distinction from the gyromagnetic Rayleigh-vorticity channel lies in the microscopic origin and the prefactor, rather than in the presence or absence of a zero-temperature term.

The corresponding Rayleigh-wave estimate follows by replacing the bulk strain field in Eq.~(\ref{eq:SM_H_g_modulation}) by the surface-mode strain.  Since a Rayleigh mode has $u_{ij}^{R}\sim q[\hbar/(2\rho A c_R)]^{1/2}e^{-\eta qd}$ at depth $d$ and a two-dimensional density of states, its zero-temperature rate has the parametric form
\begin{align}
\Gamma_{g,R}^{(0)}
\sim
\frac{\hbar\omega_0^5}{\rho g^2 c_R^5}
G_{\rm eff}^2\,
{\cal F}_g(\nu,\hat{\boldsymbol B})
\exp\!\left(-\frac{d}{\ell_R}\right).
\label{eq:SM_Rayleigh_g_scaling}
\end{align}
Here $G_{\rm eff}$ is the relevant dimensionless strain derivative of the $g$ tensor and ${\cal F}_g$ is a dimensionless angular and Rayleigh-profile factor.  Equation~(\ref{eq:SM_Rayleigh_g_scaling}) explains why the $g$-modulation mechanism can resemble the present prediction at the level of power counting: after using the resonance condition $\mu_BB=\hbar\omega_0/g$, both direct channels scale as $\omega_0^5$ and acquire the same Rayleigh evanescence.  However, the $g$-modulation channel carries the additional material-specific factor $G_{\rm eff}^2/g^2$, whereas the vorticity channel is fixed by the spin-vorticity coupling and by the Rayleigh-vorticity geometric factor derived in Sec.~\ref{subsec:SM_Rayleigh_vorticity}.

A further cancellation occurs in materials where the large $g$ factor itself originates from interband mixing.  The same $k\cdot p$ renormalization that changes the Zeeman factor renormalizes the spin-vorticity coupling~\cite{Matsuo2013Renormalization}.  Writing the static band correction as $\delta g_{\rm band}=g-g_0$ with $g_0=2$, the effective spin-vorticity term is
\begin{align}
H_{\rm sv}^{\rm ren}
=
-(1+\delta g_{\rm band})\frac{\hbar}{2}
\boldsymbol{\sigma}\cdot\boldsymbol{\Omega}.
\label{eq:SM_renormalized_sv}
\end{align}
For a fixed transition frequency, $|g|\mu_B B\simeq\hbar\omega_0$, while a resonant Rayleigh wave has $\Omega_R\sim\omega_0\epsilon$ up to profile factors.  Hence the matrix-element ratio of the strain-induced $g$-modulation channel to the renormalized spin-vorticity channel scales as
\begin{align}
\frac{M_g}{M_{\rm sv}^{\rm ren}}
\sim
\frac{G_{\rm eff}}{g(1+\delta g_{\rm band})},
\qquad
\frac{\Gamma_g}{\Gamma_{\rm sv}^{\rm ren}}
\sim
\left[
\frac{G_{\rm eff}}{g(1+\delta g_{\rm band})}
\right]^2,
\label{eq:SM_g_mod_vs_ren_sv_ratio}
\end{align}
again up to dimensionless Rayleigh-profile and angular factors.  Thus a large band-renormalized $g$ factor does not by itself make the strain-induced $g$-modulation relaxation competitive: the same band mixing enhances the spin-vorticity channel in the denominator.

A numerical comparison makes the distinction explicit.  Using diamond elastic parameters and an NV-scale frequency $\omega_0/2\pi=2.87~\mathrm{GHz}$, Eq.~(\ref{eq:SM_bulk_g_rate}) yields the estimate
\begin{align}
\Gamma_g^{(0)}
\simeq
9.4\times10^{-10}
\left(\frac{\partial g_\perp}{\partial\epsilon}\right)^2
~\mathrm{s^{-1}}.
\label{eq:SM_NV_g_rate_num}
\end{align}
At $3~\mathrm{GHz}$ this coefficient changes only by the factor $(3/2.87)^5\simeq1.25$, giving
\begin{align}
\Gamma_g^{(0)}(3~\mathrm{GHz})
\simeq
1.2\times10^{-9}
\left(\frac{\partial g_\perp}{\partial\epsilon}\right)^2
~\mathrm{s^{-1}}.
\label{eq:SM_NV_g_rate_3GHz}
\end{align}
Comparing this with the Rayleigh-vorticity benchmark in Eq.~(\ref{eq:SM_NV_Gamma0_num}) yields
\begin{align}
\frac{\Gamma_g^{(0)}}{\Gamma_0}
\simeq
2.6\times10^{-2}
\left(\frac{\partial g_\perp}{\partial\epsilon}\right)^2 .
\label{eq:SM_g_to_vorticity_ratio}
\end{align}
Thus even the optimistic order-one choice $\partial g_\perp/\partial\epsilon\sim1$ yields a rate of order $10^{-9}~\mathrm{s^{-1}}$.  In the same benchmark the Rayleigh-vorticity relaxation rate is larger by about $1/(2.6\times10^{-2})\simeq38$, consistent with the estimate that the vorticity contribution exceeds the $g$-modulation contribution by roughly fortyfold when the unknown strain derivative is taken at its maximal order-one scale.  The conclusion becomes even stronger for a hypothetical giant-$g$ material if the large strain response is tied to the same interband physics.  For InSb parameters with $g\simeq -49$, one has $\delta g_{\rm band}\simeq -51$~\cite{Matsuo2013Renormalization}; even taking the large estimate $G_{\rm eff}\simeq -50$ yields
\begin{align}
\left|\frac{M_g}{M_{\rm sv}^{\rm ren}}\right|
\sim
\frac{50}{49\times 50}
\simeq
2\times10^{-2},
\qquad
\frac{\Gamma_g}{\Gamma_{\rm sv}^{\rm ren}}
\sim
4\times10^{-4},
\label{eq:SM_giant_g_ratio}
\end{align}
apart from geometry factors.  The $g$-modulation mechanism is therefore not excluded.  Even a deliberately conservative giant-$g$ estimate leaves it as a small correction to the spin-vorticity channel, and its absolute zero-point rate is likely to be hidden below extrinsic low-temperature relaxation backgrounds in near-surface spin experiments.

This comparison clarifies the experimental motivation.  Strain-induced $g$-factor modulation is a legitimate direct one-phonon spin-lattice relaxation mechanism, especially in systems with appreciable spin-orbit-induced $g$-tensor anisotropy.  It is nevertheless a material-specific magnetoelastic channel.  By contrast, the spin-vorticity channel considered here is a gyromagnetic response to local lattice rotation and does not rely on a large magnetic anisotropy.  Coherent surface-acoustic-wave experiments have already observed spin-vorticity effects in metallic and magnetic systems through SAW-generated vorticity~\cite{Kobayashi2017,kurimune2020Observation,tateno2020Electrical,tateno2021Einstein}.  The present calculation asks for the single-spin, zero-point counterpart of that gyromagnetic coupling near a free surface.

\bibliographystyle{apsrev4-2}

\end{document}